# Enhanced vectors for top-k document retrieval in Question Answering

Mohammed Hammad

# ABSTRACT

Modern day applications, especially information retrieval webapps that involve "search" as their use cases are gradually moving towards "answering" modules. Conversational chatbots which have been proved to be more engaging to users, use Question Answering as their core. Since, precise answering is computationally expensive, several approaches have been developed to prefetch the most relevant documents/passages from the database that contain the answer. We propose a different approach that retrieves the evidence documents efficiently and accurately, making sure that the relevant document for a given user query is not missed. We do so by assigning each document (or passage in our case), a unique identifier and using them to create dense vectors which can be efficiently indexed. More precisely, we use the identifier to predict randomly sampled context window words of the relevant question corresponding to the passage along with the words of passage itself. This naturally embeds the passage identifier into the vector space in such a way that the embedding is closer to the question without compromising he information content. This approach enables efficient creation of real-time query vectors in ~4 milliseconds.

# TABLE OF CONTENTS

# LIST OF TABLES

# LIST OF FIGURES

# LIST OF ABBREVIATIONS

BEIR . . . . . . . . Benchmarking IR

BERT . . . . . . . Bidirectional Encoder Representations from Transformers

CMS . . . . . . . . Content Management Systems

ColBERT . . . . Contextualised late-interaction BERT

CW . . . . . . . . . CuratedWeb dataset

EM . . . . . . . . . Exact Match

ICT . . . . . . . . . Inverse Cloze Task

IR . . . . . . . . . . Information Retrieval

LoTTE . . . . . . Long-Tail Topic-stratified Evaluation (for IR)

NQ . . . . . . . . . Natural Questions

ORQA . . . . . . .Open-Retrieval Question Answering

PQ. . . . . . . . . . Product Quantization

PID . . . . . . . . . Passage ID

P . . . . . . . . . . . Passage

PV-DM . . . . . .Distributed Memory Model of Paragraph Vectors

QA . . . . . . . . . Question Answering

QID . . . . . . . . Question ID

Q . . . . . . . . . . Question

REALM . . . . . Retrieval-Augmented Language Modeling

RGS . . . . . . . . Relevance-Guided Supervision

SQuAD . . . . . .Stanford Question Answering Dataset

SOTA . . . . . . . State of the Art

TREC . . . . . . . Text Retrieval Conference (dataset)

TQ . . . . . . . . . .TriviaQuestions dataset

WQ . . . . . . . . . WebQuestions dataset

# 1 CHAPTER
# INTRODUCTION

Question Answering (QA) Systems are a field of NLP specifically information retrieval. The system aims to automatically answer human posed-questions in natural language. The answer could be a single word or a phrase.

## 1.1 Background of the Study

With recent advancements in computational power, several end to end QA systems have been developed. This task of QA is different from traditional search where the objective is to find relevant matching items.

There are broadly 2 kinds of QA systems depending on the type of information source and answering architecture.

### 1.1.1 Open domain

Open domain QA is more generalized and aims to answer a question in natural language. This is usually run on large set of unstructured documents. For example, retrieving an answer from a corpus of news articles. The articles may contain information in the form of natural text, tabular data, frequently asked questions, user comments, etc. Note that, even the information in the natural text is actually unstructured information, since it doesn't follow a format and can relate to any topic as well. Embedding all the information into the QA model is a challenging task in contrast to Closed Domain QA, owing to the unstructured property.

### 1.1.2 Closed Domain

Closed-domain QA usually involves questions in a specific domain e.g. medicine. They are built upon an existing structured knowledge base. For instance, an automated diagnosis and medication prescribing QA system, where the question and the answer both are structured, hence the model's architecture is simpler relatively. More precisely, the system involves an question intent detection module that outputs the query intent from one of the "closed set of existing intents". The prescription is then prescribed based on the intent. Note that in this case, the system is capable to handle limited set of questions, where the intent already exists. This is in contrast to Open Domain QA, that can handle general questions as well. Also, the user query in our example, can be in the form of a set of inputs apart from text. These inputs are structured and can be multiple choice prompts, checkboxes, etc.

### 1.1.3 Motivation

QA systems form bases of content discovery in several industries e.g. Education, customer support, etc. Most of the conversational interactions and chatbots use QA system as their core. Our research focuses on open domain QA. But in order to fetch the exact answer, it requires a key step of scanning all the documents in the database which is computationally expensive. This challenge gives rise to devise a retriever component to prefetch relevant documents containing the answer. This is why a QA system particularly of open domain requires a retriever component which is efficient, after which the computationally heavy operation of answer extraction can be applied. We discuss contemporary techniques developed till date and propose an efficient architecture which maps each passage to a unique identifier and uses it to incorporate retrieval capability by training with relevant question pairs. We extend our work by zero-shot evaluation, i.e testing the model on a different dataset, which was not used to train the model.

## 1.2 Problem Statement

Finding answers from passage against a query requires great computation power. The state of the art models for this task of reading comprehension are built with millions of parameters. This task becomes practically impossible when the answer is to be found from a large corpus of documents. In this work, we aim to efficiently retrieve relevant passages out of a large collection of documents/passages for a given question. We do so by creating dense vectors. Our work focuses on retrievers which are a prerequisite step before reading comprehension. The retriever pre-fetch top-k passages which are then fed to comprehension models. We present an efficient technique to retrieve relevant documents, which forms a key phase of QA. We use lesser computations while training and inference time. We achieve the retriever model by contextual training.

## 1.3 Aim and Objectives

Aim- Useful evidence (passages/documents) for a question normally discusses entities, relations from the question and events. Moreover, the evidence also includes additional information (which is a part of answer) that is *not* present in the question itself (Lee et al., 2019). Given a question and a large database, we aim to retrieve the most relevant passage from the database that contains the answer to that question.

Our research aims to develop a different approach in QA retrieval in terms of following points:

1. To create dense vectors for passages with customized embedding creation methods.
2. To create these vectors efficiently such that they take least time and space during training and real-time question embedding generation.
3. To compare performance of our retriever vectors against the vectors developed using state of the art transfer learning and knowledge distillation.
4. To quantify the trade-off/impact of not using negatives in retrieval since most of the existing works (Karpukhin et al., 2020) use supervision for negatives.

## 1.4    Research Questions

Do retriever training approaches involving Supervision, transfer learning and knowledge distillation significantly impact top-k retrieval accuracy as compared to our passage vectors approach ?

## 1.5    Significance of the Study

Modern Content Management Systems (CMS) based organizations (where content is updated frequently) are blocked by the costly computation power required to compute doc/passage vector representations using a large Reader model, and eventually compromise answer discovery by using BM25 scorer in retriever phase (which doesn't consider semantics). Hence, this scope of better answer discovery and performance efficiency motivates to perform this experiment .

Our problem statement adds QA perspective to the traditional/language-model based search. It fetches the "answer docs" instead of "related/similar docs" for a user query We extend our focus on the problem of answerability with different datasets and architecture as compared to existing works. More prominently, our work gives researchers an insight into the impact of passage identifier based vectors approach of retrieval. It studies the trade-off between retrieval rates vs latency (cost) . It suggests merging our approach with SOTA approaches to raise retrieval rates even higher.

For instance, one of the State-of-the-Art (SOTA) DPR system uses a model with millions of parameters to fetch top-1000 passages before feeding into the comprehension model. It takes ~9 hours on 8GPUS to process 21-million passages. Each real-time query being highly resource intensive and hence practically costly, is highly dependent on GPU and results in high latency to encode the query. This leads to the motivation of reducing latency at a minimal cost of top-k retrieval rate.

## 1.6   Scope of the Study

Question answering is done in two phases retriever and reading comprehension. The retriever phase scans the whole corpus against the user query, while the comprehension model finds the exact answer using question and retriever passages as input. As mentioned earlier, the retriever phase is a direct result of computationally expensive comprehension models. We limit the scope of this study to the retriever module. Given the time and resource constraint, we will also sample a certain percentage of data points from the already mentioned datasets. We evaluate our model on the multiple datasets and also report zero-shot evaluation metrics to study model generalization.

## 1.7   Structure of Study

In this work, we start with introducing the problem in chapter 1. We describe the motivation and significance of the efficient retrievers, and elaborate the objectives. We described in detail the significance of open-domain QA.

In Chapter 2, we review and compare the state of the art work done on retrievers. We start the chapter with a basic retriever and then describe the architecture of a recent dense passage retriever. We then study different characteristics of retriever model specifically supervision setups, attention, query-document interactions types, pre-training techniques, compression mechanism, outer-loop retrieval, performance and evaluation metrics. We also briefly present the datasets and evaluation sets extensively used for retriever training.

In chapter 3, we first present the data pre-processing we require in order to train our proposed model. Next, we discuss in detail our proposed model along with evaluation of the model and its generalization. We discuss the key intuition behind our model. We then describe the experimental configurations for training. We conclude this chapter by discussing the resource requirements required by our model to train.

In Chapter 4, we perform the ablation study for the configurations of different architectures on one of the datasets. We present the metrics and optimal hyper-parameters achieved after fine-tuning, and analyze them. We put forward the the reasons behind working of a particular choice of hyper-parameters. We then extend the optimal experiment to the remaining datasets and present the metrics for them. Lastly, we also present the generalized evaluation metrics f our model.

Next, in Chapter 5, we compare and discuss the results obtained for different datasets. We focus on top-k retrieval rates and mean rank as evaluation metrics. We also present the model latencies in a granular way.

In Chapter 6, we conclude by focusing on novelty and contribution of our work towards existing knowledge. We put forward the relevant future work and recommendations at last.

# 2 CHAPTER
# LITERATURE REVIEW

In this chapter, we analysis most recent approaches used for training retrievers. We take into account different perspectives.

## 2.1 Introduction

Contemporary Question Answering (QA) systems work in two phases or tasks. 1) Retriever to fetch top-k documents from the database of millions of docs. 2) Reader to extract exact answers spans from retrieved docs.. We go though different characteristics and perspectives for the retrievers that have been used till date.

## 2.2 Traditional Retrievers

Initial QA implementations used tf-idf (or BM25 ) as retrievers as in DrQA (Chen et al., 2017). The authors merged a bigram hashing-based search component with a TF-IDF matching component. As TF-IDF weighted bag-of-word vectors, articles and questions were compared. After that, using n-gram characteristics, particularly bigram counts, to care for local word order, the bigrams are then mapped to 224 bins using an unsigned murmur3 hash. With any query, the Document Retriever returns 5 Wikipedia articles. Then Document Reader processes the articles.

The datasets used are SQuad, CuratedTREC, WebQuestions and WikiMovies. Since, the latter 3 datasets only contain question-answer pairs, a distant supervision procedure is employed to associate paragraphs to these training examples, after which these examples are added to the training set. The authors further train the Document Reader using multitask learning by jointly training on the SQuAD dataset and all the distant supervision sources. Such a retrieval matches exact keywords instead of semantics leading to a reduced end-to-end Exact Match (EM) score compared to contemporary systems. In contrast to the DPR (Karpukhin et al., 2020) discussed in the next section, DrQA doesn't use any deep learning mechanism.

## 2.3 A brief look into a simple yet efficient retriever

Most recently, the task of retrieval is seen as a metric learning problem (Karpukhin et al., 2020) where the goal is to create an embedding-space in such a way that relevant question-document pairs are closer than irrelevant ones. The authors aim to establish if a better embedding model be trained using only pairs of questions and passages, without requiring any additional pre-training. The model leverages pretrained BERT (Kenton et al., 2019) , as an embedding function, and is based on a dual encoder architecture. The objective is to optimize the embedding by maximising the inner products of question and passage vectors, while comparing all question-passages pairs in a batch.

Two independent encoders map questions and passages to a d-dimensional (d=768) real valued vector space respectively and all the passages are indexed.

Encoders are trained in this way such that the dot-product similarity turns into a useful ranking function for retrieval. Each instance in training data comprises of a question, a relevant (positive) passage, along with n irrelevant (negative) passages. The loss function is defined as the negative log likelihood of the positive passage. Negative passages are chosen using three approaches:

1) Gold- here the positive passages are combined with additional questions from the training set.

2) BM25- BM25 returned top passages that do not have the answer even though they match the most of the question words

3) Random- here any arbitrary passage is taken.

Their top model combines one BM25 negative passage with gold passages from the same mini-batch. This approach of in-batch negatives also makes the training efficient via dynamic programming.

At run time, the question encoder maps the question to its embedding, which is then used to retrieve top k closest passages. The comparison similarity between the passage and the question vectors is calculated by dot-product.

### 2.3.1 Document Source

Wikipedia dump (English) from December 20, 2018 is used as the document source for question answering. There pre-processing is done using DrQA (Chen et al., 2017) , which breaks the paragraphs into 100 word passages after removing tables and lists. The title of the Wikipedia article is appended at the beginning of each passage as well.

### 2.3.2 Selection of positive passages

The highest-ranking passage from BM25 which contains the solution was selected as the positive passage because TREC, WebQuestions, and TriviaQA6 only offer pairs of questions and answers. If the answer is not found in any of the first 100 passages, the question was eliminated. Given that the original passages for SQuAD and Natural Questions had been split and processed in a different way than the current pool of candidate passages, each gold passage was matched and replaced with the appropriate passage present in the pool of candidates. This used to automatically associate paragraphs to such training examples, and then add these examplesse questions were discarded for which the matching has failed due to different Wikipedia versions or preprocessing.

For each dataset a model is trained, for 40 iterations for large datasets(SQuAD, NaturalQuestions, TriviaQA) and for 100 iterations for small (WQ , TREC) with Adam optimizer and batch size of 128. For better generalization, a model is also trained for all the datasets excluding squad.

## 2.4 Supervision Mechanisms in Retrieval

In this section we go through different approaches used for obtaining and/or extending the dataset for retrieval task.

### 2.4.1 No Supervision

(Lee et al., 2019) uses vector representations and yield an improvement over BM25 match.
The authors suggest Open-Retrieval Question Answering (ORQA), which does not rely on a black-box information retrieval (IR) system to find potential evidence candidates or on string supervision of the supporting evidence. Instead, they demonstrate that it is possible to jointly learn the retriever and reader end-to-end using just question-answer string pairs and without the use of an IR system. Wikipedia evidence retrieval is considered as a latent variable, and an Inverse Cloze Task (ICT) is used to pretrain the retriever.

### 2.4.2 Relevance Guided Supervision (Weak)

As compared to ORQA (Lee et al., 2019) work, several recent retrievers have shown improvement using weak supervision, one of them being ColBERT. The scalable ColBERT (Khattab and Zaharia, 2020) retriever for open-domain QA is extended by ColBERT-QA (Khattab et al., 2020) . The authors introduce Relevance-Guided Supervision (RGS), an iterative weak supervision technique that effectively samples accurate positives and challenging negatives for training. The RGS strategy enables the retriever to control its own training in order to overcome significant tradeoffs in supervision, such as expecting "positive" passages that are manually labeled- this kind of passages might not exist for all the cases; using simple models for training for instance, the BM25 technique to select "positives" and "negatives", which may yield positives that are weak and uninformative negatives; conducting within the training-loop retrieval, that necessitates often reindexing a big corpus (for example, several iterations in REALM (Guu et al., 2020) ) or an encoder that freezes the documents which makes it unadaptable to the task of retrieval (e.g. RAG; (Lewis et al., 2020)).

Instead of the expensive pretraining, RGS begins by gathering the top-k passages for each training question using an existing weak retrieval model (such as BM25). A weak heuristic is used after that, which sorts these passages into sets of negative and positive examples, based on the retriever's ordering. The heuristic is to assume that any paragraphs that include the brief answer string are positive. The previous retriever model serves as the foundation for this retrieve-and-filter process during each training round.

### 2.5 Attention Based Retrievers

Several recent systems also utilise multi-vector type representations (2.6.4). This includes Poly-encoders (Humeau et al., 2019), PreTTR (MacAvaney et al., n.d.), and MORES (Gao et al., n.d.), but they focus on re-ranking which is based on attention in contrast to scalable MaxSim-based interaction in ColBERT (Khattab and Zaharia, 2020).

## 2.6 Query-passage interaction

In this section, we study several neural matching paradigms that have been used in retrievers. Neural matching paradigms define the input internal structure of the model w.r.t to the inputs which are query and the passage. We specifically study four broad ways of query passage interaction and describe the pros and cons of the final type of interaction.

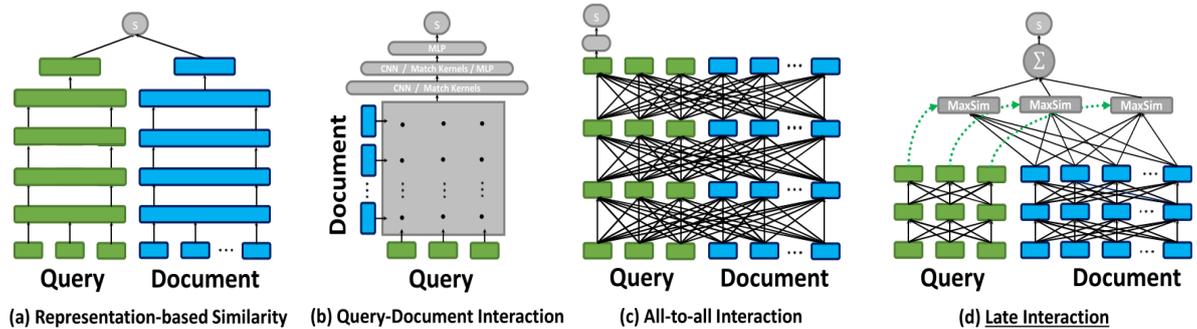

Figure 2.1. Neural Matching Paradigms (Khattab and Zaharia, 2020)

### 2.6.1 Representation-Based Rankers

Considering relevance as a similarity score between two vectors, representation-based rankers independently calculate embedding for the question, q and for the document d (Zamani et al., 2018) . Representation-Based similarity interaction is illustrated in Figure 2.1 a. Note that this architecture is different from the one mentioned in the following sub section in terms of query document interaction, which inherently doesn't happen for Representation-based rankers.

### 2.6.2 Query-Document Interaction-based Rankers

Instead of summarising the question and the document into individual embeddings, these type of rankers use deep neural networks to match phrase and word-level correlations between q and d. (attention mechanism). The interaction matrix indicates how similar all word pairs in q and d are to one another (Xiong et al., 2017) . Query-Document Interaction-based Rankers is depicted in Figure 2.1 b. This technique extends (Zamani et al., 2018) , by introducing the query-document interaction.

### 2.6.3 All-to-All Interaction Based Rankers

(Nogueira and Cho, n.d.) presents a more robust interaction-based paradigm that simultaneously represents word interactions both within and across q and d, as in BERT's architecture. Figure 2.1 c shows the All-to-All interaction mechanism. Notice the difference in architecture of such rankers with respect to Late Interaction mechanism in Figure 2.1 d.

### 2.6.4 Late Interaction (Multi-Vector Representation)

ColBERT (Khattab and Zaharia, 2020) used a different strategy by introducing a contextualised late interaction operator that independently uses BERT to encode the query and the document, then a fast yet effective interaction step to model their fine-grained similarity. It greatly speeds up query processing by utilising the expressiveness of deep Language Models while also obtaining the capability of offline pre-computation of document representations.

In the case of late interaction, the document d and the query q are independently encoded into two sets of contextual embeddings, and relevance is assessed using computations between the two sets that are favourable to pruning. The quick computations allow ranking without thoroughly analysing each potential candidate. Figure 2.1 depicts the architecture of Late Interaction mechanism.

Similar to ColBERT, COIL (Gao et al., n.d.) generates document embeddings at word-level, but the only interactions of the words are lexical matches between document and query terms. By limiting the token embedding vector of COIL to a single dimension and replacing the vector weights with scalar weights, the uniCOIL (Lin et al., n.d.) model extends methods like DeepCT (Dai and Callan, 2019) and DeepImpact (Mallia et al., 2021).

Similar to BERT, the models SPLADE (Formal et al., 2021a) and SPLADEv2 (Formal et al., 2021b) provide a vocabulary-level vector which is sparse and preserves the word-level decomposition of the late interaction simultaneously condensing the space into a single dimension per word. However, these models directly build upon BERT's pretraining language modelling capacity. SPLADEv2 has shown to be quite successful both within and across domains.

### 2.6.5 Discussion

While interaction-based models (2.5.2 and 2.5.3) often perform better at IR tasks (Guo et al., 2019), a representation-based model isolates computations among q and enables pre-compilation of document representations offline, which significantly reduces the computational load per query. The query–document interaction are retained but delayed to integrate the fine-grained matching of the interaction-based models (which imparts a great retrieval relevance ) and the document representation of representation-based models' pre-computation (which imparts retrieval efficiency) .

The MaxSim operator, which calculates maximum similarity (for instance., cosine similarity), connects each query embedding to every document embedding. These operators' scalar output are added together for all query terms.

## 2.7    Pre-Training Mechanisms

Several retrievers can be differentiated based in pre-training mechanisms them employ. We now go through several of such mechanisms used till date.

### 2.7.1    Inverse Cloze Task (ICT)

In the retriever (Lee et al., 2019) , an Inverse Cloze Task is used as pretraining. The objective of the standard Cloze (Taylor, 1953) task is to predict the material which is masked-out. The prediction is done using its context. ICT needs to predict the reverse instead; given a sentence, determine its context. The retrieval score is defined as the dot-product of vector embeddings of the query q and the passage block p. This makes the retriever learnable.

### 2.7.2    Masked Language Modeling

In contrast to the ICT task, (Guu et al., 2020) uses additional training task to achieve better results. They demonstrate for the first time how to pretrain such a generative knowledge retriever in an unsupervised manner using backpropagation through a retrieval phase that takes into account millions of documents and masked language modelling as the learning signal. Retrieval-Augmented Language Model (REALM central)'s tenet is to get the retriever trained using a signal from the unsupervised text which is performance-based: a retrieval that increases the perplexity of the LM and is rewarded is rewarded, while a retrieval that is uninformative is penalized. By using a latent variable LM and a retrieve-then-predict strategy, along with marginal likelihood optimization, this behaviour is obtained.

During the pre-training step, this model is not encoding the knowledge inside it, but it's learning to search for the right document from the external repository. For computational efficiency, since the retriever considers millions of candidate documents for each pretraining step, back-propagate through its decisions, caching and asynchronous updates are done and selection of the best documents is formulated as Maximum Inner Product Search (MIPS).

For both unsupervised pre-training for LM masking and supervised fine-tuning for downstream open-QA, REALM learns a distribution p(y | x) over possible outputs y given an input x. Masked language modelling is the task for pre-training; The model should predict the value of the tokens masked off, y, given a sentence from a pre-training corpus X with a few words masked out . Hence, the task is fine-tuning  for Open-QA for question, x, and answer y.

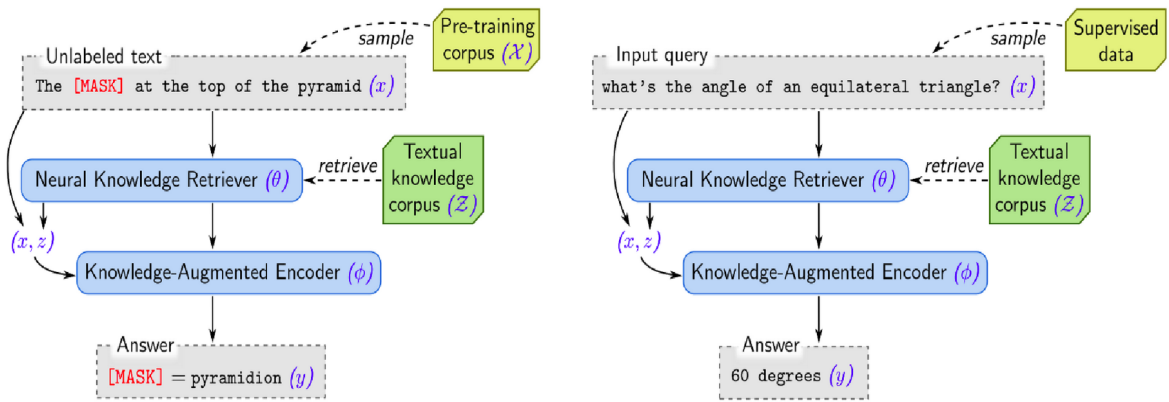

Figure 2.2. REALM Framework. Left: Unsupervised LM. Right: Supervised downstream task (Guu et al., 2020)

### 2.7.3   Other Techniques
 (Xiong et al., 2020) used high quality data generated from existing sequence model followed by progressive pre-training for negatives. This is different from ICT or Masked Language Modelling pre-training which doesn't focus on negatives.

### 2.8   Retriever Model Compression
We briefly touch upon models that improved quality using compression techniques. As compared to the ones already discussed, these techniques not only enhance latency performance but accuracy as well, since they employ a different learning design.

### 2.8.1 Distillation

(Santhanam et al., 2021) extends the work done on retrieval by using aggressive residual-compression mechanism along with denoising supervision approach. This enhances both the late interaction's quality and its spatial footprint. To improve quality beyond what is currently possible, this late-interaction type of retriever uses a combination of hard-negative mining and distillation from a cross-encoder.

The authors hypothesize that the base model, ColBERT(Khattab and Zaharia, 2020) possess a "lightweight" semantic space: not requiring the need for any separate re-training, the vectors for each interpretation of a token would cluster together, with just a small amount of context-related change. Each token's embeddings in the vocabulary are localized to only a few areas in the vector embedding space, which correspond to the contextualised "senses" of each word in the vocabulary. The cross-encoder scores are compressed into the ColBERT architecture using a KL Divergence loss.

For single-vector retrievers, (Izacard et al., n.d.) investigated dimension reduction, product quantization, and passage filtering. (Yang and Seo, 2020) emphasis on important of an independent retrieval module and further enhances end-to-end EM compared to (Hinton et al., 2015). Both use distilling of knowledge from reader to retriever. (Luan et al., 2021) compares performance of different architectures and proposes the most efficient one.

### 2.8.2 Hashing

Contrary to the distillation techniques mentioned in the previous section, (Yamada et al., n.d.) presents Binary Passage Retriever (BPR) in which embeddings are hashed directly to binary codes. This learning is done using a differentiable tanh function.

### 2.8.3 Product Quantization (PQ)

Using product quantization to compress embeddings, JPQ (Zhan et al., 2021) and its extension, RepCONC (Zhan et al., n.d.), simultaneously train the question encoder and the centroids created by Product Quantization using loss functions which are ranking-oriented. Note that these techniques differ from (Yamada et al., n.d.) in terms of archtecture design of the compression.

### 2.8.4 Autoencoder-Based

The contextual embeddings utilized for attention-based re-ranking are first reduced in dimension by an autoencoder before being further compressed using a quantization approach by SDR (Cohen et al., n.d.) .

## 2.9 Inner loop retrieval, Retrieve-and-filter and Outer loop Retrieval

In (Khattab et al., 2020), the fundamental premise is as follows; more successful retrieval will gather precise and varied positives in addition to negatives which are more difficult and realistic. RGS completely captures positive and negative feedback exterior relative to the training loop. Because of this, positives and negatives can be drawn from deep samples, and the model as a whole can be fine-tuned without often re-indexing the corpus.

### 2.9.1 A Challenge and its solution

Repeating this approach on the training set increases the chance of overfitting, which rewards a small number of positives and drifts in the negatives it samples. In order to prevent this from happening, the training set is split deterministically so that any of the model does not see a single query during retrieval and training.

Typically, during training, OpenQA, that is weakly supervised (Lee et al., 2019; Guu et al., 2020) provides passages as evidence by its own. It does this by using the short answer string to a question as a key supervision signal. A passage that contains the answer to a question in the training set, q, is considered a possible candidate for a positive passage. Two techniques—retrieve-and-filter and inner-loop retrieval—are employed to screen out erroneous matches.

### 2.9.2 Retrieve-and-filter

To find suitable passages for training, retrieve-and-filter essentially intersects a retriever that is already existing and with a weak heuristic. E.g. (Karpukhin et al., 2020) . The heuristic is such that the answer-string is more dependable signal. The candidates for q are considered from the passages that are highly ranked using BM25 are considered for q as candidates. Such methods demonstrate simplistic term-matching biases of BM25.

### 2.9.3 Inner-loop retrieval

In this architecture, the model that in the fine-tuning phase, uses the training to recover passages for each instance using the model that is being fine-tuned in inner-loop retrieval. Moving towards retriever training which is "end-to-end", this is carried out by ORQA, REALM, and RAG. However, for each training batch , as it's impossible to apply forward and backward passes on the complete set, such inner loop retrieval necessitates significant approximations. Here, while fine-tuning for OpenQA, RAG, REALM, ORQA, and freeze the document encoder along with the vectors that are indexed, are freezed, which limits the model's capacity to adapt to this job and/or to new corpora.

At the time of the document encoder fine-tuning and re-indexing a just few more times, RGS provides efficient and scalable approach in which the trainng samples are self-gathered by the retriever. By using outer-loop retrieval, RGS eliminates these restrictions.

## 2.10 Retriever Model Architectures and Prediction

The retrieval score is defined as the dot-product of vector embeddings of the query q and the passage block p, as has already been mentioned, in order for the retriever presented by (Lee et al., 2019) to be learnable. Dot products of encoders that are based on BERT are used to determine retrieval scores during runtime. The question and the evidence blocks with top scores are encoded jointly, and an MLP scores the span representations. The final joint model score is retrieval score + MLP (multi-layer perceptron) score.

$P(y \mid x)$ is broken down by REALM (Guu et al., 2020) into two steps: retrieve and predict. First, potentially helpful documents z are pulled from a knowledge corpus Z given an input x. This is modeled as sample taken from distribution $p(z \mid x)$. Next, in order to get the output y, which is represented as $p(y \mid z, x)$, the recovered-z and initial input-x are conditioned. z is used as a latent variable. All the documents, z that are possible, are marginalised over to determine the overall likelihood of generating y. $p(z \mid x)$ is modelled by the the neural knowledge retriever, and $p(y \mid z, x)$ is modelled by the knowledge-augmented encoder. Both are the two essential components.

The embedding functions are implemented using BERT-style Transformers. A projection (linear) is used to lower the vector dimensionality. The softmax function is computed over all the relevance scores. It represents the retrieval distribution. When z and x are combined prior to predicting y at the encoder, z and x are applied with rich cross-attention.

In (Khattab and Zaharia, 2020), the query encoder encodes the given query q into fixed-size embeddings' bag, and the document encoder encodes the given document d into different bag. More importantly, each vector embeddings in both the bags is contextualized w.r.t the other words in q or d, respectively. ColBERT calculates the relevance score between q and d by late interaction using both sets of bags. The maximum cosine similarity is found for each embedding in the set of query vectors with document set vectors, and the outputs are combined using summation. The intuition is, that the interaction approach, "softly" searches every query token in a way which takes into account its query context w.r.t the embeddings of the document, and hence, quantifies the "match" strength by looking for query term with the highest similarity score to a document term. It then calculates the document relevance based on these term scores by using the summation of the matched evidence w.r.t all query tokens.

ColBERT is differentiable end-to-end. There are no trainable parameters in the interaction mechanism. For a triplet (query, positive document, negative document), utilizing softmax cross-entropy loss pairwise across the calculated scores of the positive document, ColBERT individually generates each document's score.

### 2.10.1 Denoised training

Recent research has suggested that denoised training using hard negatives can be used to close the gap between interaction-based and single-vector models. Each vector v in ColBERTv2 (Santhanam et al., 2021) is encoded as an index of the centroid that is closest- C and a quantized vector r. The residual is approximated as $r = v - C$. The residual r and the centroid index recover an approximation of $v = C + r$ at the time of search. Additionally, to encode r, each of its dimensions is quantized into one or two bits while maintaining model quality for a variety of jobs downstream, greatly reducing storage costs.

### 2.11 Datasets

We use Wikipedia dump of Dec. 20, 2018 as the document source similar to (Karpukhin et al., 2020). For question passage pair we use following datasets:

## 2.11.1 Natural Questions (Kwiatkowski et al., 2019)

The dataset contains questions that were mined from actual Google search queries. Moreover, the answers were wikipedia article spans which were explicilty identified by annotators. This was the first large publicly available data set to pair real user queries with high-quality annotations of answers in documents.

A dataset of questions and answers, the Natural Questions (NQ) corpus includes 307,373 training instances, 7,830 development examples, and 7,842 test examples. Google.com search terms and corresponding Wikipedia pages make up each example. Each Wikipedia article has one or more short spans from the annotated passage that contain the actual answer as well as a passage (or lengthier answer) that is annotated on the page that answers the inquiry. As a result, each data point is presented as a triple comprising a question, wikipedia page, long answer, and short answer. However, both the long and short answer annotations may be empty. There is no answer on the page at all if they are both empty. The annotated paragraph answers the questions if the long response annotation is non-empty and the short answer annotation is empty. However, no explicit short answer could be found. Finally, instead of a list of brief spans, 1% of the documents have a passage annotated with a brief "yes" or "no" answer.

Every question in NatualQuestions (NQ) is an 8-word or longer search query that has been submitted to Google over a brief period of time by numerous users. The question asks for factual facts; the Wikipedia page could or might not include the data needed to respond.

A pool of about 50 annotators worked on the annotation. The quality of the question is determined by contributors. A good question is one that is fact-seeking and can have an entity or explanation as a response. Bad questions are ambiguous, difficult to understand, based on blatantly erroneous assumptions, opinion-seeking, or not explicitly a request for factual information. Annotators are not yet shown the Wikipedia page, so they must judge the question entirely on its own merits.

In all, annotators identify long answers for 49% of the examples, short answers for 35%, yes/no answers for 1%, long answers for only 13% of the examples, and no answers for 37%. The remaining 51% of the questions that have no labelled answers are not discarded because the decision of whether or not to answer a question is viewed as a vital component of the question-answering task.

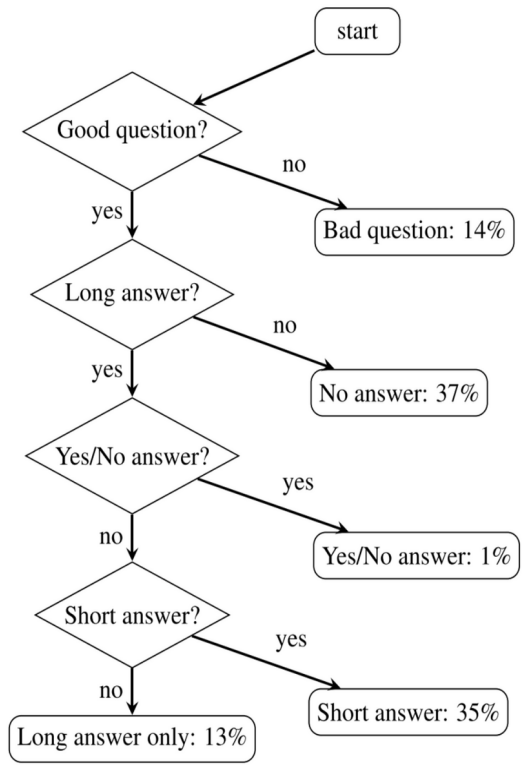

Figure 2.4. Annotation Decision process for NQ (Kwiatkowski et al., 2019)

Figure 2.3. NQ dataset instances (Kwiatkowski et al., 2019)

### 2.11.2 TriviaQA (Joshi et al., 2017)

TriviaQA (TQ) contains web scrapped question and answers. 950K question-answer pairings from 662K pages culled from Wikipedia and the internet make up this realistic question-answer dataset. Given the length of the context and the possibility that the answers to a question cannot be determined through span prediction, this dataset is more difficult than traditional QA benchmark datasets like the Stanford Question Answering Dataset (SQuAD).

The TriviaQA dataset includes QA subsets that were created automatically and by humans. The dataset creation process comprises evidence documents that are independently acquired with average of 6 per qquery; that gives excellent distant supervision for question answering. The questions in the dataset are rather intricate and sophisticated in nature. Between questions and the related answer-evidence sentences, there is a significant amount of syntactic and lexical variation. Finding solutions calls for more cross-sentence reasoning.

| Property | Example annotation | Statistics |
| --- | --- | --- |
| Avg. entities / question | Which politician won the **Nobel Peace Prize** in 2009? | 1.77 per question |
| Fine grained answer type | What **fragrant essential oil** is obtained from Damask Rose? | 73.5% of questions |
| Coarse grained answer type | **Who** won the Nobel Peace Prize in 2009? | 15.5% of questions |
| Time frame | What was photographed for the first time in **October 1959** | 34% of questions |
| Comparisons | What is the appropriate name of the **largest** type of frog? | 9% of questions |

Figure 2.5. TriviaQA properties (Joshi et al., 2017)

The first step was to collect pairings of question-answer from fourteen websites of trivia or quiz league. Since they were typically either too simple or too unclear, questions with fewer than four tokens were eliminated. Two sources were used to gather textual evidence for the answers: The top 50 search result URLs are collected, and the automatically obtained documents are from the Wikipedia or generic results of Web search.

> **Question**: The Dodecanese Campaign of WWII that was an attempt by the Allied forces to capture islands in the Aegean Sea was the inspiration for which acclaimed 1961 commando film?
> **Answer**: The Guns of Navarone
> **Excerpt**: The Dodecanese Campaign of World War II was an attempt by Allied forces to capture the Italian-held Dodecanese islands in the Aegean Sea following the surrender of Italy in September 1943, and use them as bases against the German-controlled Balkans. The failed campaign, and in particular the Battle of Leros, inspired the 1957 novel **The Guns of Navarone** and the successful 1961 movie of the same name.
>
> **Question**: American Callan Pinckney's eponymously named system became a best-selling (1980s-2000s) book/video franchise in what genre?
> **Answer**: Fitness
> **Excerpt**: Callan Pinckney was an American fitness professional. She achieved unprecedented success with her Callanetics exercises. Her 9 books all became international best-sellers and the video series that followed went on to sell over 6 million copies. Pinckney's first video release "Callanetics: 10 Years Younger In 10 Hours" outsold every other **fitness** video in the US.

Figure 2.6. TriviaQA instances (Joshi et al., 2017)

The evidence was acquired using an automated procedure, thus it cannot be assured that it has all the facts required to respond to the inquiry. In light of the presumption that the inclusion of the answer string in a piece of evidence means that the piece of evidence actually provides an answer to the query, they are best viewed as a source of distant supervision.

| Reasoning | Lexical variation (synonym) |
|---|---|
| | Major correspondences between the question and the answer sentence are synonyms. |
| Frequency | 41% in Wiki documents, 39% in web documents. |
| Examples | Q What is solid CO2 commonly called? |
| | S The frozen solid form of CO2, known as **dry ice** ... |
| | Q Who wrote the novel The Eagle Has landed? |
| | S The Eagle Has Landed is a book by British writer **Jack Higgins** |
| Reasoning | Lexical variation and world knowledge |
| | Major correspondences between the question and the document require common sense or external knowledge. |
| Frequency | 17% in Wiki documents, 17% in web documents. |
| Examples | Q What is the first name of Madame Bovary in Flaubert's 1856 novel? |
| | S Madame Bovary (1856) is the French writer Gustave Flaubert's debut novel. The story focuses on a doctor's wife, **Emma** Bovary |
| | Q Who was the female member of the 1980's pop music duo, Eurythmics? |
| | S Eurythmics were a British music duo consisting of members **Annie Lennox** and David A. Stewart. |
| Reasoning | Syntactic Variation |
| | After the question is paraphrased into declarative form, its syntactic dependency structure does not match that of the answer sentence |
| Frequency | 69% in Wiki documents, 65% in web documents. |
| Examples | Q In which country did the Battle of El Alamein take place? |
| | S The 1942 Battle of El Alamein in **Egypt** was actually two pivotal battles of World War II |
| | Q Whom was Ronald Reagan referring to when he uttered the famous phrase evil empire in a 1983 speech? |
| | S The phrase evil empire was first applied to the **Soviet Union** in 1983 by U.S. President Ronald Reagan. |
| Reasoning | Multiple sentences |
| | Requires reasoning over multiple sentences. |
| Frequency | 40% in Wiki documents, 35% in web documents. |
| Examples | Q Name the Greek Mythological hero who killed the gorgon Medusa. |
| | S **Perseus** asks god to aid him. So the goddess Athena and Hermes helps him out to kill Medusa. |
| | Q Who starred in and directed the 1993 film A Bronx Tale? |
| | S **Robert De Niro** To Make His Broadway Directorial Debut With A Bronx Tale: The Musical. The actor starred and directed the 1993 film. |
| Reasoning | Lists, Table |
| | Answer found in tables or lists |
| Frequency | 7% in web documents. |
| Examples | Q In Moh's Scale of hardness, Talc is at number 1, but what is number 2? |
| | Q What is the collective name for a group of hawks or falcons? |

Figure 2.7. Reasoning analysis of TriviaQA (Joshi et al., 2017)

### 2.11.3 WebQuestions (Berant et al., 2013)

There are 6,642 question-answer pairings in the WebQuestions (WQ) dataset, a question-answer dataset that uses Freebase as the knowledge base. It was built by crawling queries through the Google Suggest API and then utilising Amazon Mechanical Turk to get answers. The original division employs 2,032 instances for testing and 3,778 examples for training. Every answer has been defined as a Freebase entity.

The questions are mostly centered around a single named entity. The questions are popular ones asked on the web. The authors used a breadth-first search across questions (nodes), starting with the query "Where was Barack Obama born?" and using the Google Suggest API to provide the edges of the network in order to retrieve the questions. To be more precise, they ran queries on the question excluding the entity, the phrase before the entity, and the phrase after it; each query produced five candidate questions, which were then put to the queue. A random 100K questions were sent to Amazon Mechanical Turk after they iterated till 1 million questions had been visited (AMT). Workers were instructed to respond to the query using only the Freebase page for the entity it related to, or else label it as unanswerable by Freebase, according to the AMT job. Out of 100K questions, 6,642 had identical annotations from at least two AMT employees.

### 2.11.4 CuratedWeb (Baudiš and Šedivý, 2015)

A curated and expanded version of the TREC corpus makes up the dataset. The TREC-8 (1999) to TREC-13 (2004) Question Answering tracks made up the original Text Retrieval Conference Question Answering v (TrecQA) dataset. The benchmarks from the TREC QA tasks serve as the foundation for the CuratedWeb (CW) dataset. The 2001 and 2002 TREC QA tracks were carefully examined, questions that were deemed irrelevant or out-of-date were eliminated, and the patterns were updated using recent statistics or Wikipedia terms. It has 867 open-domain factoids divided into 430 training (and development) questions and 430 testing questions at random. The question-answer combinations came from actual user queries sent to search engines like MSNSearch and AskJeeves.

### 2.11.5 Squad v1.1 (Rajpurkar et al., 2016)

A span of text from the corresponding reading passage serves as the answer to each question in the Stanford Question Answering Dataset (SQuAD), a reading comprehension dataset made up of questions posed by crowdworkers on a collection of Wikipedia articles. The question itself may also be unanswerable. It is more diversified than some other question-answering datasets since the questions and answers are generated by people through crowdsourcing. SQuAD 1.1 has 536 articles with 107,785 question-answer pairings.

This dataset is useful for reader training and not retriever since many questions lack context. We include it since several recent works have been trained upon it as well. Two more datasets were used for experimentation, used by ColBERT (Khattab and Zaharia, 2020) are mentioned in the next subsequent section.

### 2.11.6 MS MARCO

Introduced by Microsoft in 2016, MS MARCO is a dataset for reading comprehension and was enhanced for retrieval in 2018. It is a collection of 8.8 million Web page passages that were taken from Bing's responses to 1 million actual queries. Each query is linked to sparse relevance judgments of one (or a few) documents labelled as relevant and no explicitly designated irrelevant documents.

### 2.11.7 TREC CAR.

Introduced by (Dietz et al., 2012), TREC CAR is a synthetic dataset based on Wikipedia, contains around 29 million passages. The title of a Wikipedia page and the heading of one of its sections were combined to create 3M queries. The passages in that section that are relevant to the corresponding query are marked as relevant.

## 2.12 Evaluation datasets

We now briefly present some recent, retrieval oriented evaluation datasets. These datasets involve several retrieval tasks for evaluation.

### 2.12.1 BEIR (Thakur et al., 2021)

A variety of information retrieval tasks are included in the BEIR (Benchmarking IR) benchmark. The zero-shot generalisation abilities of various neural retrieval techniques can be thoroughly studied with BEIR. The benchmark includes nine information retrieval tasks from 17 different datasets, including fact checking, citation prediction, duplicate question retrieval, argument retrieval, news retrieval, question answering, tweet retrieval, and entity retrieval.

### 2.12.2 LoTTE

Additionally, the authors presented a new benchmark called LoTTE, which stands for Long-Tail Topic-stratified Evaluation for IR. It consists of 12 domain-specific search tests that span StackExchange communities and use questions from the GooAQ (Khashabi et al., n.d.) dataset. Unlike the Open-QA tests and many of the BEIR tasks, LoTTE concentrates on long-tail topics in its passages and evaluates models on their ability to respond to natural search questions with a practical intent, as opposed to many of the BEIR's semantic-similarity tasks.

## 2.13 Retriever performances and learnings

In situations, where the queries depict an information need, that is, when the query authors don't know the answer already, (Lee et al., 2019) establishes that learning to retrieve is essential. Since, there is intrinsic biases in all the known datasets, which is troublesome for open domain QA models with learnt retrieval, the authors analyse on 5 ORQA datasets. The questioners do not already know the solution in the CuratedTre, WebQuestions, and Natural Questions. This is an accurate representation of the distribution of real inquiries. ICT's pre-trained retriever fared significantly better than BM25 on these questions because they were taken from actual users who are looking for information. Annotators required assistance from automatic tools while finding correct answers which introduced just a moderate bias. Contrarily, since the questions in TriviaQA and SQuAD are constructed with predetermined answers in mind, automatic tools are not required. Since writing the question is not driven by any information need, it significantly skewed the dataset that was obtained. Therefore, ORQA does not outperform BM25 retrieval.

The REALM model (Guu et al., 2020) was evaluated on 3 datasets: NaturalQuestions, WebQuestions (Berant et al., 2013)and CuratedTREC. The model outperformed the previous benchmark of ORQA for all 3 datasets and established the importance of pre-training. This model with only 300 million parameters performs as well as 11 billion parameters used in the earlier T5 language model. It's much easier to debug where it's getting it's answers from, as the model can now provide with an index of the document it referred, to support the arguments it's making. REALM had the overall greatest performance despite retrieving only 5 documents, as opposed to other retrieval-based systems that frequently retrieve between 20 and 80 documents.

BM25 and DPR (Karpukhin et al., 2020) based two top-2000 passage sets are obtained first, respectively, then reranked the union of them using linear weight of 1.1. On every dataset, DPR consistently outperforms BM25. TREC, which is the tiniest of the five datasets, gains considerably from extended training instances when training with multiple datasets. TriviaQA marginally deteriorates, while Natural Questions and WebQuestions make only moderate improvements.

### 2.13.1 Reason for low accuracy on SQuAD

There are two causes behind the SQuAD's decreased performance. After reading the passage, the annotators first wrote questions. Hence, there's significant amount of lexical overlap between the texts and the questions, which clearly favours BM25. Second, the distribution of training instances is highly skewed because the data was only gathered from 500+ Wikipedia articles.

DPR is trained exclusively on Natural Questions and tested directly on the smaller CuratedTREC and WebQuestions datasets in order to demonstrate the generalisation across datasets. It was discovered that DPR generalises well, surpassing the BM25 baseline significantly (55.0/70.9) with just a 3-5 point loss from the top-performing fine-tuned model in top-20 retrieval accuracy (69.9/86.3 vs. 75.0/89.1 for WebQuestions and TREC, respectively).

The dense vector representations so trained outperformd Lee et al., 2019 in Top-5 accuracy (65.2% vs. 42.9%) . The authors demonstrate that BM25 can be significantly outperformed by merely tweaking the question and passage encoders on pre-existing question-passage pairings. Furthermore, they demonstrate that in the case of open-domain question answering, a greater retrieval precision does actually translate to a higher end-to-end QA accuracy by establishing a higher end-to-end QA accuracy on Natural Questions (41.5 % vs. 33.3 %).

Run-time efficiency for DPR (Karpukhin et al., 2020) : Passage retrieval speed was profiled on intel Xeon CPU E5-2698 v4@ 2.2Ghz with 512MB memory. The DPR model processes 995 questions/second, returning top 100 passages/question. Computing dense embeddings on 21M passages for indexing in the initial stage took 8.8 hours on 8 GPUs.

In contrast, the authors of ColBERT (Khattab and Zaharia, 2020) establish that, in comparison to conventional BERT-based models, it is more than 170 times faster and uses 14,000 times less FLOPs per query while beating non-BERT based models. In the subsequent version, the authors show that ColBERT-QA (Khattab et al., 2020) outperformed other models (within and outside the training domain), particularly on the three the datasets;  NaturalQuestions, TriviaQuestions and SQuaD.

### 2.13.2 Zero-shot evaluation of ColBERTv2 (Santhanam et al., 2021) on IR tasks

ColBERTv2 had the highest MRR@10 of any standalone retriever when trained on MS MARCO Passage Ranking. Three Wikipedia Open-QA tests and 13 various retrieval and semantic similarity tasks from BEIR (Thakur et al., 2021) are among the several benchmarks used to assess ColBERTv2. Using its compressed representations on 22 of the 28 out-of-domain tests, ColBERTv2 earned the highest quality, surpassing the next best retriever by up to 8% relative gain.

## 2.14 Summary

We discussed some traditional retrievers based on bag of words. Then we studied the architecture of a recent retriever which guided in understanding the basic idea of retriever mechanism. We then presented different types of supervision approaches used in contemporary retrievers. This was followed by attention based retrievers. We then dived into interaction mechanisms and discussed their architectures along with contemporary implementations. Importantly, several pretraining mechanisms were discussed, along with their hypothesis. Next, we studied compression techniques used in retrievers. We then focused more on architecture of several retrievers particularly, the denoised training. Lastly, we went through the training and evaluation datasets used by contemporary retrievers and their performance.

# 3 CHAPTER
# RESEARCH METHODOLOGY

In this chapter, we deep dive into dataset preparation and preprocessing. We then discuss in detail the model architecture and the experimental setup.

## 3.1 Introduction

The datasets available for retriever training were annotated directly on the web pages. Hence, they are wrapped in HTML tags and contain noise. We discuss the preprocessing done on the datasets and the Wikipedia dump, which acts as the source/corpus for information retrieval. Our proposed model trains the question passage pair by assigning them a unique identifier and then embedding into into vector space. This is discussed in more detail in the Proposed Model Section 3.3. We experiment with different setups to achieve the best retriever model. We train model for each daaset separately and a combined model. Lastly, we evaluate our models per dataset and for datasets that were not used for training.

## 3.2 Data Preprocessing

We extract the text from the passages of the Natural Questions dataset since they contain HTML tags. The passages are then cleaned and tokenized using the same scripts of DPR (Karpukhin et al., 2020) . The purpose of using the scripts is unbiased comparison of the results. We thus obtain question-passage pairs.

As mentioned earlier, English dump of Wikipedia dated from Dec. 20, 2018 is used as source of documents. This is also preprocessed by removing HTML tags using the same scripts DrQA (Chen et al., 2017) .After cleaning, the text-portion of articles is extracted. This step Any semi-structured data, for instance, lists, tables, info-boxes, disambiguation pages are removed in this step. Each article is then split into multiple, disjoint text blocks, each of 100 words. These passages form basic retrieval units. The title of the Wikipedia article is prepended to each passage along with an [SEP] token.

For all the question-passage pairs obtained from the NQ dataset, each passage is checked for existence in the pre-processed Wikipedia article. This is done using string matching. In case the passage is not matched with any of the passages from the dump, the data instance is not considered. The remaining NQ dataset questions are then split into training, validation and test set in the ratio 80/10/10 .

Similarly, the steps are repeated for other datasets as well namely, TriviaQA, WebQuestions, CuratedWeb and MS MARCO.

## 3.3 Proposed model

In our approach, each passage and question are assigned a unique id, respectively.

### 3.3.1 Architecture

Each passage id is mapped to an embedding in vector space of dimension, n. This passage embedding is randomly initialized. Given a question and the corresponding passage containing the answer evidence, each training instance consists of the set of question tokens and the passage id. The passage id is shared for all the sub training instances generted by the moving window. The data samples generated are trained in an *unsupervised* way, such that the question tokens and the passage id gets embedded into vector space relevantly. More precisely, each token in the set is predicted using the set of neighbouring tokens with a window size, k and the passage id. This is done for all question passage pairs in the dataset.

In order to embed maximum information from passages to their respective embeddings, similar approach is used for passage id and passage tokens. Specifically, the passage tokens in the moving window along with the shared passage id across these tokens are used to predict the main token one after the other. This is done as a different configuration experiment and is discussed in more detail in the experimentation section below.

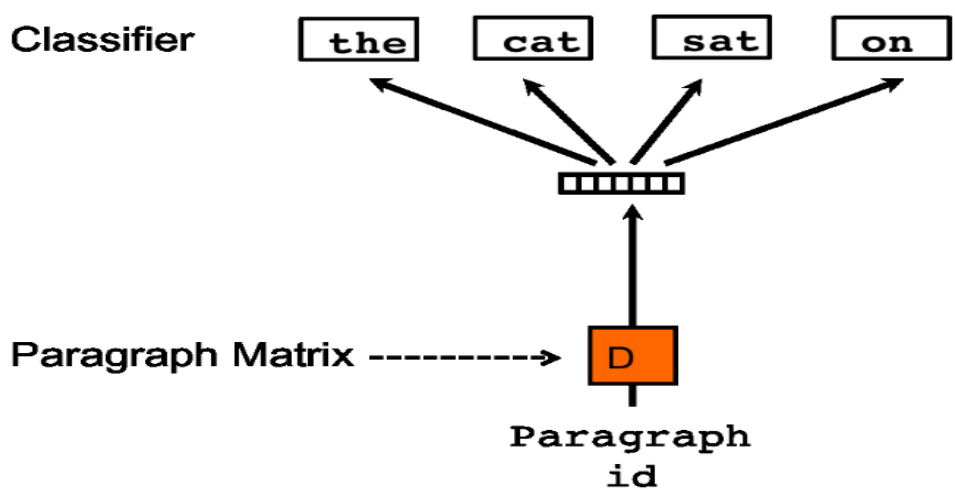

Figure 3.1: Model architecture to train passage embeddings (Le and Mikolov, 2014)

### 3.3.2 Runtime Inference

The architecture discussed trains a model which is capable of generating question embeddings for a question at realtime. The by-product of the model training is the set of passage embeddings. These embeddings are indexed (using FAISS, ElasticSearch) in a database for efficient matching. For a question, q and a passage,p, we define our similarity score, sim($q, p$) as the cosine distance between the question and the passage embeddings. This scores each passage against the question, and the passage are ranked and retrieved.

$$\text{sim}(q, p) = \cos(Embed_Q(q), Embed_P(p)) \quad (1)$$

where Embed are the vectors for question and passage.

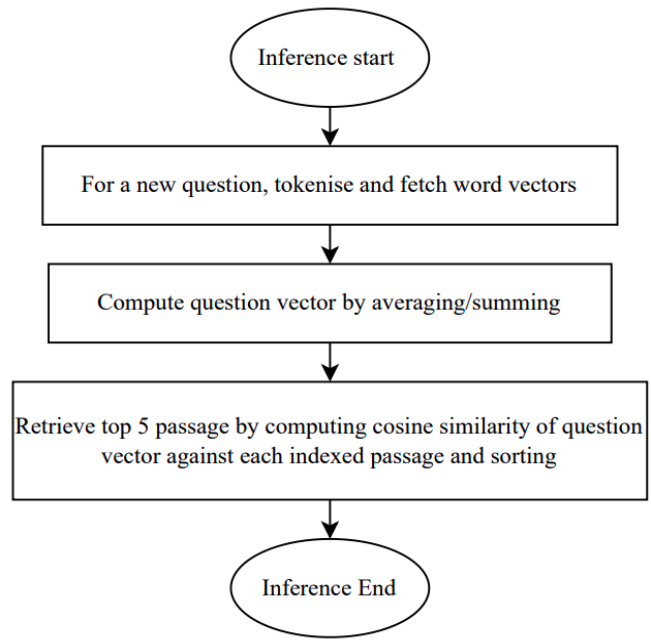

Figure 3.2. Model Realtime prediction flow

### 3.3.3 Key Intuition

In order to create efficient passage-vector representations capable for retrieving relevant passages, our model is extended by (Le and Mikolov, 2014) . Our approach is motivated by the techniques of learning the word vectors. The motivation is that the word vectors are made to contribute to the task of prediction of each word in the sentence. So despite randomly initializing the word vectors, they are able to eventually capture the relevant semantics. This is an indirect result of the prediction task.

We hypothesis that when the passage vectors are also made to contribute to the prediction task of the main word given moving window samples from the passage, it naturally embeds the word vectors and the passage id that are retrieval-conscious.

The passage id can be viewed as another token. It acts as a memory that remembers what is missing from the current context, or the topic of the passage. The configuration is referred as Distributed Memory Model of Paragraph Vectors (PV-DM) (Le and Mikolov, 2014).

### 3.3.4 Training

Standard gradient descent is used to optimize. We experiment with the window size of 5, 10, vector size of 100, 500, learning rate range 0.001 – 0.1. We employ the technique of Bayesian optimization in order to get the best set of parameters for our model. Figure 3.3 show the training procedure in detail. Given the words from the passage window context along with the passage identifier, PID, both as input, the obejective is to predict next word in the passage. This is done by maximizing the average of log of probability,

$$\frac{1}{T}\sum_{t=0}^{T-k} \log p(w_t | PID, QID, w_{t-k}, ....., w_{t+k}) \qquad (2)$$

We use softmax for multi-class classification,

$$p(w_t | PID, QID, w_{t-k}, ....., w_{t+k}) = \frac{e^{y_t}}{\sum_i e^{y_i}} \qquad (3)$$

Each $y_i$ is un-normalised log of probability corresponding to each token i,

$$y = b + Uh(w_t | PID, QID, w_{t-k}, ....., w_{t+k}; W) \qquad (4)$$

where U, b are softmax parameters, h is average of extracted word vectors

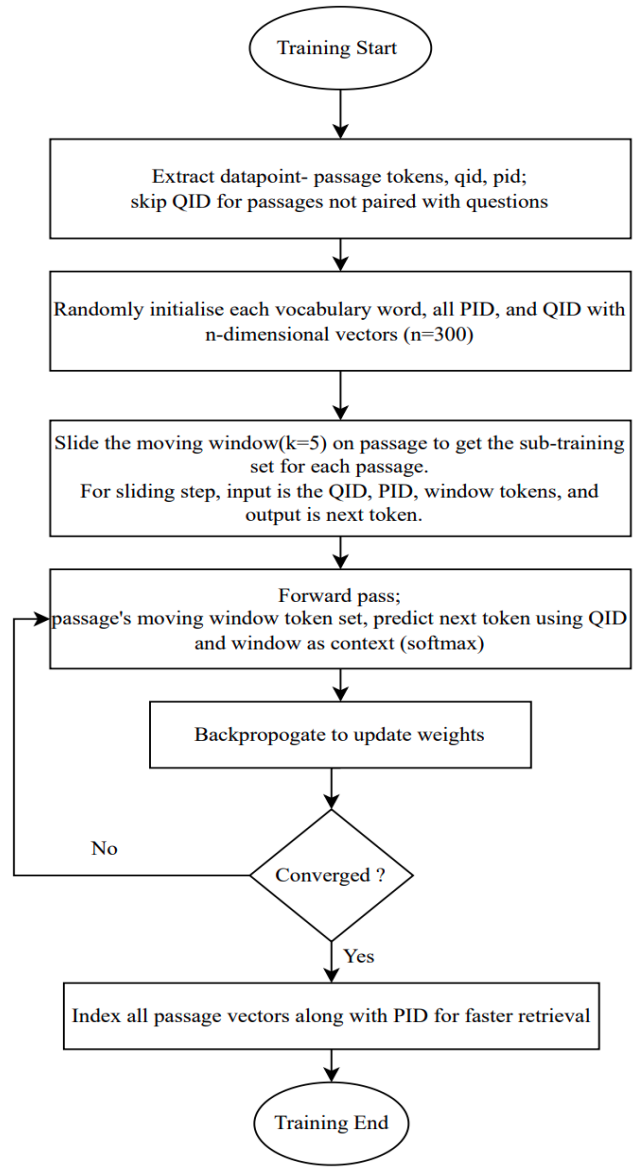

Figure 3.3. Model Training Flow

## 3.4 Experiments

We experiment with different settings : question and passage as a single document. Question as tags, two–way encoder for question answer pair trained in both directions. The reader is not trained in our case. The details for the experimental setup has been mentioned in section 4.5. For each experiment, we define our hypothesis and present the metrics after the ablation. The experiments primarily differ in terms of usage of passage, passage Id (PID), question and the question Id (QID). At each step, the next experiment configuration is decided based on the results of the previous ones. The optimal performing experiment is gradually fine-tuned to give better results.

## 3.5 Datset setups

The parameters of best model obtained from the Experiments defined in section 3.4 are kept constant for the following settings mentioned.

### 3.5.1 Single dataset model

The model is trained on each dataset separately. We use NQ, TriviaQA, CuratedWeb, WebQuestions datasets for this configuration.

### 3.5.2 Combined dataset model

Apart from training the model on each dataset independently, we also train a model on all the datasets combined. This is expected to perform best since it not only corresponds to the best parameter setting from Section 3.4 experiments, but also, has diverse sources of information. We report the performance of this model the test set of each dataset separately.

## 3.6 Evaluation

In this section, we discuss the evaluation techniques used to compare model performance and generalizability.

### 3.6.1 Evaluation Metric

In order to evaluate our approach, we use the top-k retrieval metric, k = 1 and 5. For instance, to compute top-5 retrieval, each passage is scored against the question in test set and then sorted. In case the evidence passage is there in the top 5 passages, we mark the test data point as successfully predicted, else failed. The metric is reported as the percentage of the number of data points successfully predicted out of total data points in the test set.

### 3.6.2 Zero-shot Evaluation

In order to evaluate the model generalization, the model is evaluated on a different dataset from the one it was originally trained. We evaluate the model trained on NQ dataset, on TriviaQA and compare the results with previous implementations. This helps in determining the model's transfer learning capabilities.

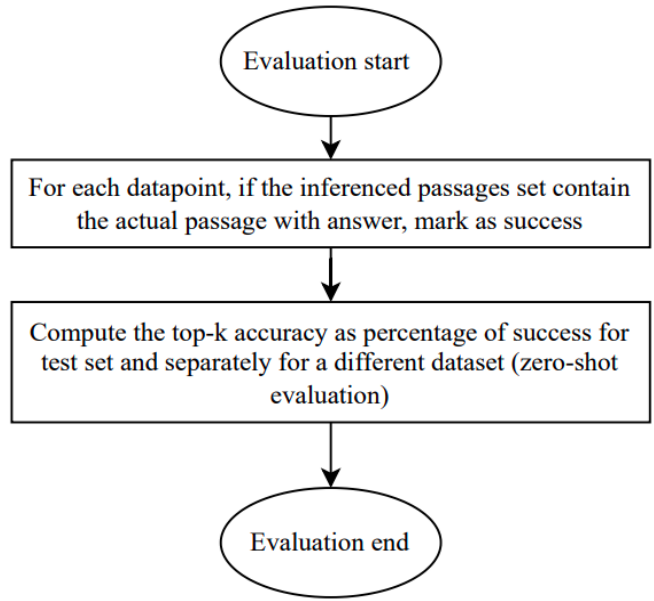

Figure 3.4. Model Evaluation flow

## 3.7 Novelty

For the first time, we use this approach of contextual word training in retrieval. The vectors obtained are efficient since the model used is a shallow. Compared to non-transfer learning approaches. Our model doesn't require any pre-training. Also, we don't use any transfer learning like the contemporary BERT systems. We have detailed out the contribution of this work to existing knowledge in the final chapter.

## 3.8 Resource Requirements

We use the following dependencies in our work

### 3.8.1 Hardware Requirements

The model is trained on A6000 GPU with 8GB GPU memory and 32GB RAM. The upper limit levels of RAM are reached during the merged dataset experiment. The machine comprise of 32 cores of CPU.

### 3.8.2 Software Dependencies

We use python3.8 in our work. We particularly used the following libraries for models training 1) gensim for experiments 1-5 , 2) multiprocessing and BayesianOptimization.

## 3.9 Summary

In this chapter we presented the data preprocessing techniques used by the retriever model. We then presented the proposed model, and discussed the key intuition behind its working. We described novelty in our approach which makes our work different from existing approaches. We systematically described our dataset setups for experimentations. We concluded with evaluation techniques and the approach to evaluate model generalizability.

# 4 CHAPTER
# ANALYSIS

In this Chapter, we describe our actual datasets used for training in terms of the corpus, training, validation and test set. We also analyze our choice of training parameters and hyper-parameters and explain the reason behind the optimal values of the parameters. All the experiments mentioned in section 3.6 are analyzed and the optimal hyper-parameter values are discussed for the experiment that gave the optimal retrieval rate.

## 4.1 Introduction

An important aspect of retrieval is the architecture of the model being trained. Training on all the corpus, without efficient model design is practically impossible due to resource constraints. Even the SOTA models use a simple heuristic model (for instance, tfidf) to fetch top n (say n=1000 ) passages, and then feed these passages into the retriever model. In this chapter, we start with exploring our dataset and our corpus. We conduct the ablation study of a number of architecture level experiments mentioned in Chapter 3. We present hyper-parameters of the optimal ablation experiment and the hypothesis behind those optimal values. Finally, we conduct the cross dataset experiment to prove our model generalization.

## 4.2 Corpus Preparation

The preprocessed wiki corpus and the datasets are available publicly [here](). The data is preprocessed using techniques as described under section 3.2. As part of tokenization step, we make sure the all the punctuation and repeating characters are handled. For instance, the phrase "during same year(1940)" is tokenised into "during", "same", "year", "(", "1940" and ")". Similarly, the repeating characters, for instance "what???" is tokenised into just two tokens instead of four. This is done for the corpus and all the datasets as well. The corpus originally contains 21015324 passages. We downsize the corpus to 976939 passages due to resource and time constraints.

## 4.3 Data Transformation

In order to feed our data into the model, we link each passage id mentioned in the question set to the actual passage id. We prepend "p_" before passage id, "nq_" , "tq_" , "wq_" and "cw_" before the question ids for the corresponding dataset. This step is important especially during the general model experiment. Every question instance finally contains the question id, and the answer passage id list. Every passage id contains the corresponding passage id and (optionally question id based on experiment). We also downsample the corpus due o resource constraints according to counts mentioned in Table 4.1.

## 4.4 Exploratory Data Analysis

The preprocessed NQ training set contains of 58880 questions corresponding to a total of 498816 passages. Since a question can have the answer in multiple passages, each question corresponds to an average of 8 passages. The maximum passages corresponding to a particular question are 101. The validation and test set contain a total of 3257 questions each. Both the sets have a total of 51251 passages out of which 24629 were not present in the corpus. Note that the training, validation and test set of passages being an evidence to a question, are actually part of corpus. Hence, while training dense vectors we create vectors for the passages of corpus (976939), training set (325652), validation set (3257) and test set (3257), which is a total of 1309106 passages. This is done for faster retrieval of passages during validation and testing.

*Table 4.1: Comparison of training, validation and test set for various dataset*

| Dataset | Corpus | Training questions | Distinct Training passages | Trainig: Corpus Passage Ratio | Validation | Test | Max. passages / question | Avg. passages / question |
|---|---|---|---|---|---|---|---|---|
| NQ | 976939 | 58880 | 325652 | 1:3 | 3257 | 3257 | 101 | 8.5 |
| TQ | 1541577 | 60413 | 513859 | 1:2 | 4418 | 4418 | 100 | 12.3 |
| WQ | 1208241 | 2474 | 23691 | 1:50 | 278 | 278 | 100 | 11.1 |
| CW | 774690 | 1125 | 15190 | 1:50 | 115 | 115 | 100 | 14.9 |

Similarly, we repeat the process for other datasets and present the summary in table 4.1. The corpus section in the table represents passages corresponding to training questions and the passages not corresponding to them but downsampled from wiki corpus. For TriviaQuestions dataset, only 60413 questions contain answer passages out of the provided 78785. Hence, we used the reduced set for this case.

## 4.5    Ablation Study (Natural Questions)

We start with analyzing results for each experiment mentioned in section 3.4 on NQ dataset. We present the metrics for all experiments to justify our choice of best experiment. Later, we proceed on other datasets with the experiment that gave optimal results. The model training was accomplished on Intel i5-1145G7 @ 2.60GHz CPU. We also provide a comparison between retrieval time as a function of corpus size.

### 4.5.1    Experiment 1 (Retrieval-oriented context)

In this configuration we consider the passage id, PID along with question,Q. We predict the question words using PID. The results have been summarized in Table 5.1. The hypothesis is that the combination of passage id with question imparts retrieval capability. The question id- question combination imparts enhanced realtime vector computation capability. Briefly, we consider:

PID-Q

### 4.5.2    Experiment 2 (Retrieval-oriented passage context)

We extend experiment by adding passage id – passage pairs to the dataset. The hypothesis is that this will help in retaining information that might not be their in questions and should perform better on zero-shot evaluation. Briefly, we consider:

PID-Q

PID-P

### 4.5.3    Experiment 3 (Retrieval-oriented question context)

In this configuration we consider question id and passage id along with Q. The hypothesis is that the combination of question id with passage may also imparts retrieval capability The question id- question combination imparts enhanced real-time vector computation capability. We summarize the results in table 5.3. Briefly, we consider:

QID-P

QID-Q

PID-P (the usage decision is based on experiment 1 and 2 accuracies)

### 4.5.4 Experiment 4 (Merged General)

For thi configuration, we consider all possible pairs. Briefly, we consider:

QID-P

PID-P

QID-Q (PV-DM)

PID-Q (PV-DM)

## 4.6 Handling Retrieval Rank for Multi-Passage Questions

We faced the challenge in the rank and retrieval metric calculation for questions with multiple passage answers. In such cases, we took the average rank of passages that were retrieved. A default max rank of 1000 was assigned corresponding to questions which retrieved no passage. Suppose a question retrieves 5 passages within max rank i.e 1000, with each passage rank as 5, 13, 27, 31, 50 respectively. The rank would be average of the 5 ranks i.e (5 + 13 + 27 + 31 + 50)/ 5 = 25.2. Note that the question may originally have 10 answer passages in the corpus. As mentioned earlier, this question with rank of 25.2 will be considered in top-100 retrieval

## 4.7 Hyperparameters Tuning

In this Experiment 2, the PID is trained with the passage text and the corresponding question text. This intrinsically makes the PID vector embed into vector space in such a way that the question and passage come closer to each other. While training, we fine-tune our model with different parameters and summarize our choice of parameters below.

### 4.7.1 Vector Size

We experiment with 100, 500, 650, 800 and 1000 size of vectors. The optimal size comes out to be 800. All the sizes below 800 gave lesser training and validation set retrieval rate metric. This is due to lesser information capability. Also, greater sizes tend to trade off generalization since information storage among the vectors becomes sparse which was evident from the significant difference between training and validation set.

### 4.7.2 Window Size

The window sizes of 3, 5 and 10 were used. 5 turned out to be optimal. The window of 3 didn't work well due to lesser context being accounted during training. On the contrary, the window of 10 was not giving meaning full results since there is too many words but little information to be learnt.

### 4.7.3 Min-count

In order to remove noise words and create vectors for words where there is atleast sufficient information, we had to remove words less then a vocabulary count of 2. We observed that increasing the threshold degraded the results.

### 4.7.4 Distributed-Memory Configuration

For our case of passage retrieval, we observed that training the model by *using the PID vector to predict words randomly sampled from the passage window context gave better results then training by predicting the word from its window context and PID*. This configuration works better since relatively more information is embedded into the PID vector, in contrast to word prediction which attributes limited information.

### 4.7.5 Learning Rate/ Epochs

We experimented with the learning rate range 0.0001 – 0.1 and minimum learning rate range of 0.0001 – 0.01 . The optimal learning rate was 0.014 while the minimum training rate varied from 0.005 – 0.01 for different datasets. The training was done for 15 epochs after which any further training resulted into over-fitting.

## 4.8 Dataset wise Experiment Analysis

In this section, we extend our experiment 2 to other datasets and present the metrics we achieve for each case. For each dataset, we start training with the same set of hyper-parameters on the other datasets, that had given optimal results on NaturalQuestions dataset.

### 4.8.1 TriviaQA Experiment

We summarize the results in Table 4.6. This dataset has slightly reduced retrieval rate compared to that of NaturalQuestions. This is due to the difference in the dataset sourcing technique. TriviaQA contains questions that are inherently more difficult to answer than NaturalQuestions.

*Table 4.2: Experiment 4 metrics on TriviaQuestions*

| TQ | Mean Rank | Top-1 | Top-10 | Top-20 | Top-100 |
|---|---|---|---|---|---|
| Training | 141 | 7.6 | 38.7 | 49.4 | 53.8 |
| Validation | 313 | 3.3 | 17.8 | 28.7 | 36.3 |
| Test | 405 | 2.7 | 13.2 | 18.3 | 30.0 |

### 4.8.2 WebQuestions Experiment

Experiments on this dataset resulted in a significant drop in both the metrics, top-20 retrieval rate and mean rank. The training set top-100 retrieval rate itself dropped from 56.9 on NaturalQuestions to 31.9 on WebQuestions. This observation arises from the fact that WebQuestions dataset has only 2474 training questions as compared to almost 59k questions in NaturalQuestions. This further translates to the validation and test set metrics as visible in Table 4.7.

*Table 4.3: Experiment 4 metrics on WebQuestions*

| WQ | Mean Rank | Top-1 | Top-10 | Top-20 | Top-100 |
|---|---|---|---|---|---|
| Training | 191 | 5.2 | 24.6 | 27.8 | 31.9 |
| Validation | 370 | 3.5 | 17.5 | 22.0 | 25.9 |
| Test | 452 | 2.1 | 10.1 | 13.2 | 17.6 |

### 4.8.3 CuratedWeb Experiment

The problem of reduced dataset instances is more visible on the CuratedWeb dataset. The dataset contains merely 1125 training questions compared to almost 59k in NauralQuestions. The model is hardly able to learn any meaningful retrieval capability in this case. The metrics are summarized in Table 4.8.

*Table 4.4: Experiment 4 metrics on CuratedWeb*

| CT | Mean Rank | Top-1 | Top-10 | Top-20 | Top-100 |
|---|---|---|---|---|---|
| Training | 207 | 3.7 | 13.1 | 16.2 | 22.6 |
| Validation | 421 | 2.5 | 11.8 | 14.6 | 17.2 |
| Test | 457 | 1.8 | 9.8 | 12.5 | 15.0 |

For the purpose of reproducibility of our research, the preprocessing and training code is publicly available [here](here).

## 4.9 Cross-dataset Generalized Model Analysis

To understand how well does our model generalize on other datasets, we used the model trained on NQ dataset on TQ dataset and evaluated it. Table 4.9 summarizes the metrics observed for such configuration. The vectors for TQ passages were generated and indexed during training on NQ dataset. As expected, we observe a slightly reduced retrieval rate and mean rank as compared to model that was originally trained on TriviaQuestions dataset itself, the metrics of which are already mentioned in Table 4.6. This configuration proves our model has good generalizing capability.

*Table 4.5: Generalized Experiment on TriviaQuestions dataset*

| TQ | Mean Rank | Top-1 | Top-10 | Top-20 | Top-100 |
|---|---|---|---|---|---|
| Training | 189 | 6.9 | 32.7 | 41.4 | 45.8 |
| Validation | 325 | 3.1 | 14.8 | 21.0 | 33.7 |
| Test | 431 | 1.9 | 11.9 | 14.6 | 26.2 |

## 4.10 Summary

We conducted ablation study for several architectures. We also observed that experiment 2 turned out to be the optimal experiment. We observed that the validation mean rank tends to saturate after 10 epochs, visible from the training and validation mean rank as a function of epochs. We present and discuss the results achieved in the next chapter.

# 5 CHAPTER
# RESULTS AND DISCUSSION

In this chapter, we present and discuss the results obtained for the ablation performed for all the experiments. We start with comparing retrieval rates among different datasets. We then compare our metrics with SOTA model metrics. Finally, we conclude with the discussion on latencies of our retriever model.

## 5.1 Introduction

We start with comparing the top-20 retrieval accuracy we achieved with one the SOTA models. An important aspect of retriever is the latency i.e the server time required to fetch top-n results. Hence, we present the latencies achieved by our model and the impact of corpus size on it. We later focus on our contribution towards existing knowledge.

## 5.2 Ablation Analysis

We analyses each experiment in this section.

### 5.2.1 Experiment 1 (Retrieval-oriented context)

We predict the question words using PID. This gave a top-40 training retrieval rate of 49 but performed poorly on the validation set. This is particularly due to the model not being trained on passage words. As visible in Table 5.1, we observe a large gap between training and test metrics. This is particularly due to shallow training of passages not part of the training questions. Hence, the vectors generated for them are not meaningful and lead to lower validation and test metrics.

Table 5.1: Experiment 1 metrics on NaturalQuestions

| NQ | Mean Rank | Top-1 | Top-10 | Top-20 | Top-100 |
|---|---|---|---|---|---|
| Training | 209 | 8.0 | 20.6 | 45.0 | 52.2 |
| Validation | 501 | 1.1 | 8.5 | 18.2 | 22.5 |
| Test | 799 | 1.0 | 3.2 | 12.5 | 15.5 |

## 5.2.2 Experiment 2 (Retrieval-oriented passage context)

This experiment turned out to be the optimal performing model. We got the optimal mean rank of 392 on test set. The top-20 retrieval achieved is 19.5% , while the top-100 retrieval rate is 57.8%. This means that given a question, the answer is 57.8% times contained in the top-100 passages retrieved by our model. The PID actually couples the question and passage and brings their vectors closer to each other since the PID is actually being used to predict the question and passage words. We summarized the results in Table 5.2.

The Figure 5.1 (a) plots top-20 training and validation retrieval rates upto 20 epochs. We observe that the top-20 validation retrieval rate tends to saturate after 10 epochs. Figure 5.1 (b) plots top-100 training and validation retrieval rates.

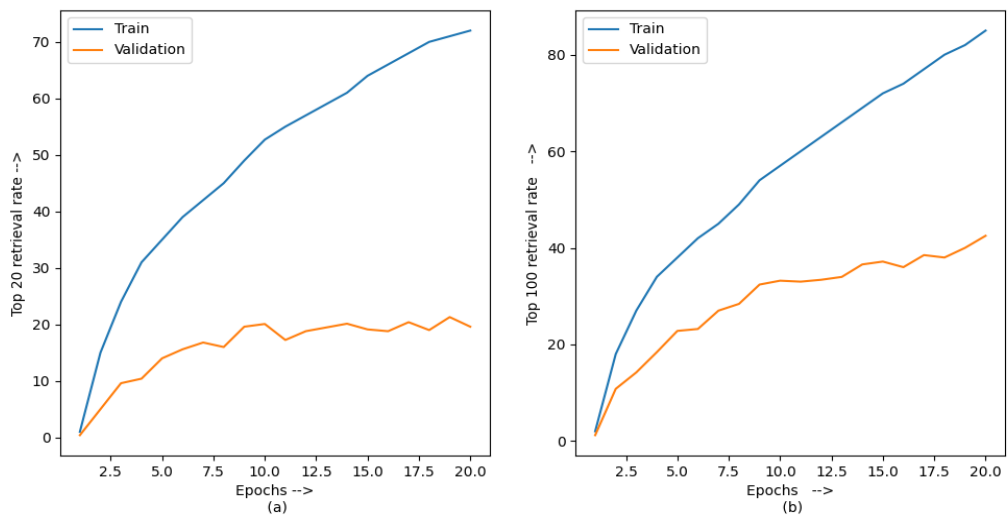

Figure 5.1: Convergence curves (a) Top-20 retrieval rates and (b) top-100 retrieval rates on NaturalQuestions

Each epoch took 900 seconds for the NQ dataset. The training times were proportional to dataset-size for other datasets. It was observed that a total of 10 epochs gave optimal results. More details about hyper-parameters are already mentioned section 4.7. In later sections, we argue that the metrics can be further improved if the model is trained on complete corpus. Table 5.2 summarizes the metrics for Experiment 2 on NaturalQuestions.

*Table 5.2: Experiment 2 metrics on NaturalQuestions*

| NQ | Mean Rank | Top-1 | Top-10 | Top-20 | Top-100 |
|---|---|---|---|---|---|
| Training | 133 | 8.0 | 40.7 | 52.7 | 57.8 |
| Validation | 300 | 3.5 | 18.5 | 29.3 | 37.4 |
| Test | 392 | 2.8 | 13.5 | 19.7 | 32.5 |

Figure 5.2 plots mean rank as a function of number of epochs upto 20. The training and validation curves also signify the convergence around 10 epochs after which the mean rank begins to saturate. This metric correlates with top-20 retrieval rate observed in Figure 5.1 a.

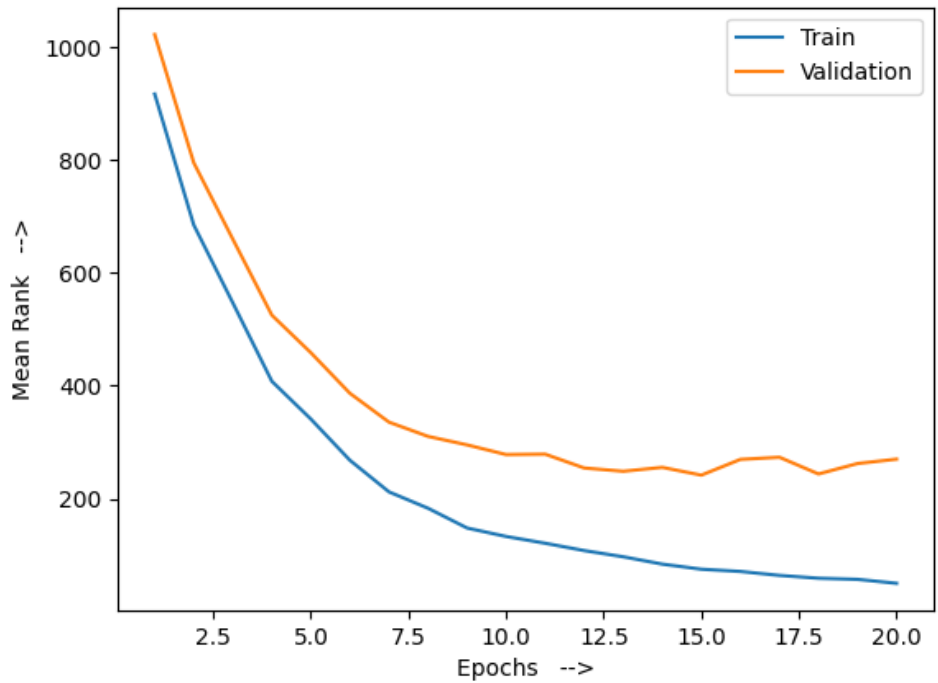

Figure 5.2: Mean Rank obtained on NaturalQuestions

### 5.2.3 Experiment 3 (Retrieval-oriented question context)

Here, instead of PID we use QID to predict the passage and question words. We also gave more weights to the question samples by further predicting question words from QID. This experiment resulted in retrieval rates that were comparable to the previous experiment in section 6.4.2. The configuration didn't improve upon the retrieval rates and mean rank achieved in Experiment 2. The results are summarized in section Table 5.3

*Table 5.3: Experiment 3 metrics on NaturalQuestions*

| NQ | Mean Rank | Top-1 | Top-10 | Top-20 | Top-100 |
|---|---|---|---|---|---|
| Training | 143 | 7.9 | 39.5 | 49.9 | 55.9 |
| Validation | 323 | 3.4 | 16.3 | 28.7 | 35.2 |
| Test | 410 | 2.7 | 12.8 | 19.3 | 30.9 |

### 5.2.4 Experiment 4 (Merged General)

We gradually built upon experiment 2 to consider the merged, question and passage words while using the PID to predict them. This gave retrieval rates comparable to Experiment 2. The training set mean rank was better then that of Experiment 2, but the retrieval rate wasn't better relatively. We summarized the results in Table 5.4.

*Table 5.4: Experiment 4 metrics on NaturalQuestions*

| NQ | Mean Rank | Top-1 | Top-10 | Top-20 | Top-100 |
|---|---|---|---|---|---|
| Training | 116 | 7.9 | 40.5 | 51.7 | 56.9 |
| Validation | 285 | 3.4 | 18.4 | 29.6 | 37.1 |
| Test | 371 | 2.7 | 13.3 | 18.6 | 31.7 |

## 5.3 Training Comparison Among Dataset

We study the comparison of retrieval rates for different datasets. Figure 5.1 shows that retrieval rate curves for all the four datasets we have trained upon separately. The impact of reduced dataset is clearly visible from the curves of WebQuestions and CuratedWeb. This also hints us to a complementary fact that a larger corpus can actually further improve our metrics.

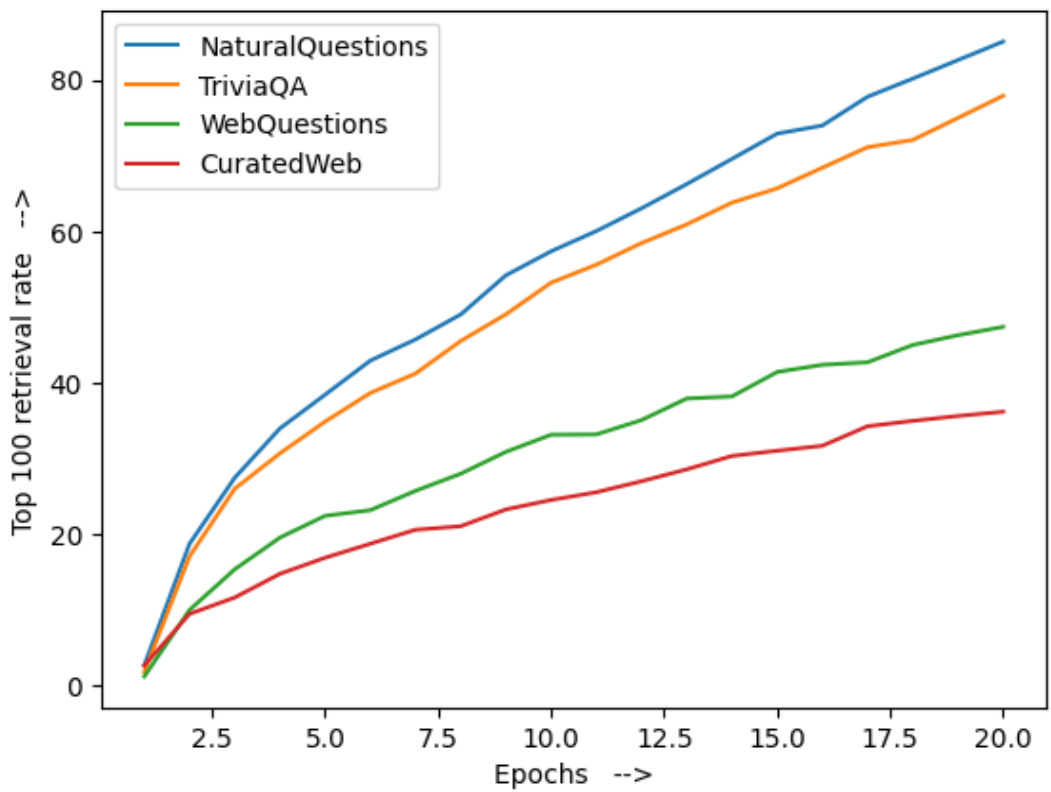

Figure 5.3: Relative comparison of multiple dataset retrieval rates

## 5.4 State-of-the-Art (SOTA) Model Comparison

Our experiments visibly appear to obtain metrics that are less effective compared to state of the art benchmarks. Table 5.1 compares one such benchmark (Karpukhin et al., 2020) for all the four datasets. For the NaturalQuestion dataset we achieve 19.7 top-20 retrieval rate as compared to 78.4 SOTA metric. This is a direct consequence of the reduced corpus, which is merely 5% of actual Wiki corpus, that we have used owing to resource constraints. This is visible from the fact that for the datasets, WebQuestions and CuratedWeb, we obtain lesser retrieval rates as they have far lesser questions relative to NaturalQuestions and TriviaWeb. Another point worth mentioning here is the fact that, we have achieved the evaluation metrics without using negative passage contexts, a perspective that several SOTA model focus upon.

*Table 5.5: Comparison with SOTA model*

| Model / Dataset | Top-20 | | | | Top-100 | | | |
|---|---|---|---|---|---|---|---|---|
| | NQ | TQ | WQ | CW | NQ | TQ | WQ | CW |
| DPR | 78.4 | 79.4 | 73.2 | 79.8 | 85.4 | 85.0 | 81.4 | 89.1 |
| Our Model | 19.7 | 18.5 | 13.1 | 11.2 | 32.5 | 31.7 | 17.3 | 13.9 |

## 5.5 Real-time Retrieval Latencies

One of the achievements this Real-time Retrieval latency consists of two major computations, the user query vector computation and the similarity match among all corpus passages. We study both the latencies separately and focus it from the point of novelty.

### 5.5.1 Query Vector Computation Latency

We observed that our query vector computation was between 3 to 4 milliseconds on Intel i5-1145G7 @ 2.60GHz CPU. This effectively removes the necessity of GPUs. This is a direct consequence of our subtle model architecture with vector size being 800. The SOTA models tend to have billions of parameters and hence take several hours just for computing passage vectors for indexing. On the other hand, the time required for building. Even parallel computation of dense embeddings on 21-million passages took roughly 8.8 hours on 8 GPUs (Karpukhin et al., 2020) . This also proves the cost effectiveness of our model.

### 5.5.2 Similarity Match Latency

The second major computation latency is indexed search to find most similar passage vectors to that of the question vector. Figure 5.2 shows latency as a function of corpus size. The curve indicates direct proportionality of retrieval latency with corpus size. Our model, for instance, took 125 milliseconds to fetch top 1000 passages form a corpus of 1327220 passages. This can be further improved using existing advanced indexing techniques (Johnson and Research Paris HervéHerv, n.d.) . This problem of Similarity match latency being already solved, we limit are our discussion here.

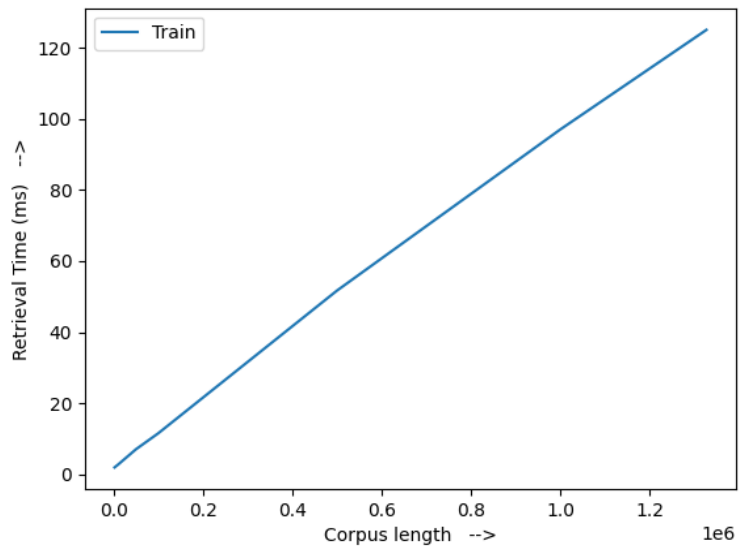

Figure 5.4: Retrieval times vs Corpus length

## 5.6 Summary

In this chapter, we discussed insights into the reason behind variation in retrieval rates among different datasets we trained our model upon. We saw how our model performs in comparison to other SOTA models. We also showed our model latency to compute the vector for the real-time user query is far less as compared to the SOTA model. The trade off is the retrieval rates. We concluded with focusing on the efficiency of our model in terms of resource requirements, particularly, on the practical fact of our model being able to compute real-time query vectors on CPU instead of GPU.

# 6 CHAPTER
# CONCLUSION AND FUTURE SCOPE

This chapter contains the main conclusions of our work. We discuss the novelty of our work in comparison to the existing work done till date. Moreover, we also discuss the future work of our study.

## 6.1 Conclusion

From the experiments we have conducted, we conclude that our neural network architectures especially the one we used in Experiment 2 in section 4.5.2, tend to be highly efficient with the trade-off on retrieval rate. Such architectures are highly impacted by the size of corpus as observed among different datasets as observed in section 4.8 and 5.2.

## 6.2 Contribution to Existing Knowledge

The novelty in our work comes from the application of the unique PID based vectors computation in Question Answering oriented passage retrieval. The practicality and cost effectiveness perspective of out work is visible from the fact that we have used a subtle model architecture that computes the 800-dimensional query vector in 3-4 milliseconds without the need for GPUs.

Our work gives researchers an insight into the impact of PID based vector approach towards retrieval. It studies the trade-off between retrieval rates vs latency (cost) . It gives a direction of combining our approach with SOTA techniques to further improve retrieval rates. We have achieved the evaluation metrics without using negative passage contexts, a perspective that several SOTA model focus upon. Similarly supervision, pre-training and other complex interaction mechanisms as mentioned in Chapter 2, have also been avoided in order to achieve efficient latencies. Finally, the metrics we achieved used merely 5% of wiki corpus.

## 6.3 Future work

It would be interesting to study the metrics and latencies by combining our work with SOTA techniques such as supervision and pre-training. Further, the impact of training on negative passage contexts with our approach should further improve our metrics. Lastly, using the corpus to its 100% potential will further yield better metrics. Owing to our resource constraints, we could utilize merely 5% of the Wiki corpus. Training on larger corpus, would enable a word be embedded into space more meaningfully since it will have several more contexts associated to it.

# APPENDIX

Dataset links:

https://github.com/facebookresearch/DPR/blob/main/dpr/data/download_data.py

Code:

https://github.com/hammad7/ms/tree/main/retrieval